\documentclass[
    ,final            
  ,numberedheadings 
  ]
  {aipproc}

\layoutstyle{8x11single}

\newcommand{\msun}{M$_{\odot}$ }

\begin{document}

\title{Blue Straggler Formation in Clusters}

\classification{97.20.Rp,97.80.-d,97.80.Kq,98.20.Gm,98.20.Di}
\keywords      {Blue stragglers; open clusters; globular clusters; binary stars; multiple stars}

\author{Alison Sills}{
  address={Department of Physics and Astronomy, McMaster University, 1280 Main Street West, Hamilton, ON, L8S 4M1}
}

\begin{abstract}

Blue stragglers are thought to be formed from the merger or
coalescence of two stars, but the details of their formation in
clusters has been difficult to disentangle. We discuss the two main
formation mechanisms for blue stragglers (stellar collisions or mass
transfer in a binary system). We then look at the additional
complications caused by the stars living in the dynamically active
environment of a star cluster. We review the recent observational and
theoretical work which addresses the question ``which mechanism dominates?''
and conclude that the most likely answer is that both mechanisms are
at work, although with different importances in different environments
and at different times in the cluster lifetime. We conclude with a
short discussion of some avenues for future work.
\end{abstract}

\maketitle


\section{search for formation mechanisms}

Blue stragglers are stars which are bluer and/or brighter than the
main sequence turnoff in clusters. They are found in all resolved
stellar populations, including globular and open
clusters, dwarf galaxies \cite{2007MNRAS.380.1127M}, and even
the field \cite{2010arXiv1002.3009D}. They are main sequence stars, lying between the zero-age main sequence and the giant branch in clusters. Typically they do not lie on a tight sequence, but instead fill much of that region of the colour-magnitude diagram. Where we can identify companions, we know that blue stragglers often occur in binary systems or even in triples
\cite{2009Natur.462.1032M, 2010arXiv1008.4347T}.  Typically there are
a few tens of blue stragglers in open clusters, and up to a few
hundred in globular clusters. While they only represent 1\% or less of
the stellar population in any cluster, they are easily identified and
provide a way to probe the history of the cluster and/or its binary
population, depending on our understanding of how this unexpected
population was formed.

Any formation mechanism for blue stragglers must find a way to either
increase the mass of a normal (low-mass) cluster main sequence star,
or delay the evolution of a higher-mass star such that it stays on the
main sequence longer than its peers. Currently, the two main formation
mechanisms which are thought to create blue stragglers are direct
stellar collisions, or mass transfer in a binary system. We discuss
each of them below, and then look at the importance of the cluster
environment.

\subsection{Stellar Collisions}

In the simplest form of this scenario, two previously unrelated stars
collide as they are moving through the cluster. A fairly dense
environment is needed, but calculations of collisional cross sections
in globular \cite{1976ApL....17...87H} and open
\cite{1989AJ.....98..217L} clusters have shown that collisions are
common enough in these environments to produce approximately the right
number of blue stragglers. In some clusters, it is expected that every
star has undergone at least one close encounter. There is also
evidence that another exotic population, cataclysmic variables, is a
collision product and they are over-abundant in clusters
\cite{2006ApJ...646L.143P}. Blue stragglers are also concentrated in
the cores of clusters \cite[e.g.][]{1992A&A...262...63L}, suggesting that
they are more massive than the average star, but also indicative that
they are formed in the densest environments.

\begin{figure}
  \includegraphics[height=.4\textheight]{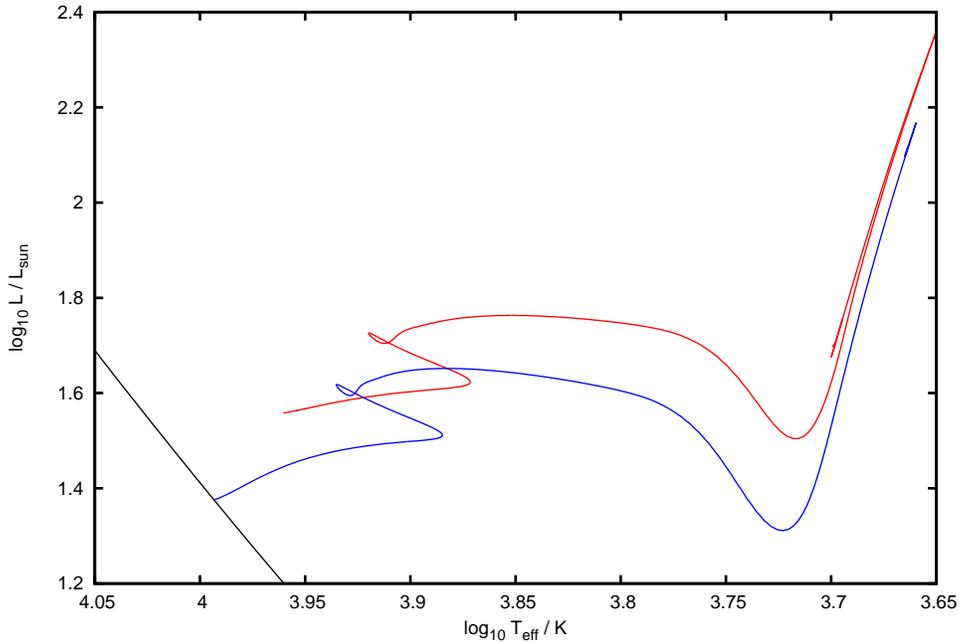}
  \caption{Evolution of the product of a collision between a 2.4 \msun
    and a 1.9 \msun star (red), compared to the evolutionary track of
    a normal 4.3 \msun star (blue). Models courtesy of E. Glebbeek,
    based on the calculations in Glebbeek \& Pols\cite{2008A&A...488.1017G}.}
  \label{EvertCollision}
\end{figure}

Hydrodynamic simulations of stellar collisions
\cite[e.g.][]{1996ApJ...468..797L} show that for the relatively gentle
collisions which occur in globular clusters, the collision product is
a star-like object. Only a few percent of the total mass of the two
parent stars is lost during the collision, and the rest settles into a
spherical, hydrostatic object. The collision product is typically not
fully mixed, but instead the cores of the parent stars end up at the
core of the collision product \cite{1996ApJ...468..797L}. Subsequent
stellar evolution calculations of the collision product
\cite{2001ApJ...548..323S, 2008A&A...488.1017G} have shown that after
some initial thermal-timescale relaxation, the star evolves relatively
normally. One such example is shown in Figure
\ref{EvertCollision}. The red line is the evolutionary track of the
product of a collision between a 2.4 and 1.9 \msun star, and the blue
line shows the evolution of a normal star of 4.3 \msun. The initial
thermal relaxation of the collision product has not been shown. In
general, collision products are quite similar to normal stars of the
same mass. However, their main sequence lifetime is generally shorter
than that of the normal star. Since the parents of the collision
product have been evolving for some time before the collision, they
have converted some of their hydrogen to helium, and so the collision
product has a smaller hydrogen reservoir and a shorter main sequence
lifetime. Their position in the colour-magnitude diagram is therefore
usually above the zero-age main sequence, reflecting their advanced
stage of evolution when they are formed. However, the collision
products do typically spend considerable time in the appropriate
region of the colour-magnitude diagram so that we would expect to see
them as blue stragglers.

\subsection{Binary Mass Transfer}

The other plausible mechanism for blue straggler formation involves
mass transfer in a binary system. We know that binaries exist in all
stellar populations, and we also know that if the orbits are small
enough, mass will be transfered from one star to another. Binary mass
transfer has been invoked to explain a wide variety of unusual stars,
starting with the Algol system and encompassing objects at all
evolutionary stages. Blue stragglers are seen in all well-studied
stellar populations, including sparse clusters and the field. Stellar
collisions are not expected to contribute to blue stragglers in these
environments, and therefore binary mass transfer provides an obvious
and likely alternative formation mechanism.

If the accreting star is a main sequence star, then the product will
be a blue straggler. Not all of the donor star needs to become part of
the blue straggler -- only enough mass to significantly change the
accretor's position in the colour-magnitude diagram needs to be
transferred. For relatively high-mass accretors near the turnoff, this
could be only a few tenths of a solar mass. The other extreme is a
mass transfer event in which the entire donor star is accreted onto
the blue straggler. Some authors refer to the latter as ``binary
coalescence'' or ``binary merger''. For the purposes of this review,
we will consider any amount of mass transfer as belonging to this
mechanism. 

The position of the blue straggler in the colour-magnitude diagram and
its main sequence lifetime depend heavily on the masses and
evolutionary states of the two stars involved. The initial binary
orbit also determines how much mass will be transfered, and when. The
parameter space of binary mass transfer blue stragglers has not yet
been fully explored. The most comprehensive studies to date follow
binaries which are appropriate for producing blue stragglers in old
open clusters like M67 \cite{2006A&A...455..247T,
  2008MNRAS.384.1263C}. Both case A and case B mass transfer have been
studied. Again, the evolutionary tracks for these objects spend a
significant amount of time in the blue straggler region of the
colour-magnitude diagram. Interestingly, many of the mass transfer systems spend
the majority of their time at or near the equal-mass binary line (0.75
magnitudes above the zero age main sequence), regardless of the actual
mass ratio of the system.

The present-day appearance of binary mass transfer products will
depend strongly on when the mass transfer began, whether the mass
transfer was conservative, and whether the mass transfer was
complete. Incomplete mass transfer will result in a blue
straggler in a binary system, and the nature of that companion star
can tell us a lot about the initial configuration. For example, a main
sequence companion means that case A mass transfer was the cause,
while a helium white dwarf points to a giant branch donor instead. 

\subsection{Influence of the cluster environment}

In the discussion above, we have considered the effects of stellar
dynamics and of binary evolution separately. In clusters, however,
this simplification is invalid. We know that all clusters have a
population of binaries, and these binaries can interact dynamically
with each other and with single stars. These interactions will modify
our expectations for the blue straggler populations in clusters in a
number of ways.

Dynamical interactions will modify the binary population in
clusters. The standard wisdom can be summarized as ``hard binaries
harden'' \cite{1975MNRAS.173..729H}. In other words, during long-range
encounters with other objects in the clusters, binaries whose orbital
energy is larger than the average kinetic energy of stars in the
cluster will have their orbital energy increased. Soft binaries, on
the other hand, have their orbital energy decreased by encounters,
even to the point of disruption of the binary completely. What
implications will this have for those binaries which would have formed
blue stragglers through mass transfer? In order to produce a blue
straggler that we identify at the current time, mass transfer needs to
have happened in the somewhat recent past.  Also, the accretor needs
to be a main sequence star of the appropriate mass. Otherwise, the
blue straggler will have had time to leave the main sequence and
evolve off to the giant branch. If the distribution of binary orbital
energies (and hence separations) is being modified by interactions,
then the number of appropriate binaries which would have undergone
mass transfer at the appropriate time will have changed. Some systems
will be hardened to the point where mass transfer will happen earlier
and we will not see them as blue stragglers at the current
time. Other, wider systems, which would not have become blue
stragglers, may be hardened enough that those objects will contribute
to the blue straggler population. However, we must also consider the
evolutionary state of the donor when determining which objects will
become blue stragglers -- the initial orbital separation is not
enough information.

In addition to long-range, gentle encounters, binaries in clusters
undergo close encounters with single stars or other binaries which
result in much more violent outcomes. Binaries can be broken apart
entirely, which means that those objects cannot later produce blue
stragglers. These close encounters are typically resonant encounters
\cite{1983ApJ...268..319H}, meaning that the individual stars spend a
long time involved in a complicated dance around each other. Exchange
interactions are quite common, and the typical outcome is that the
heavier stars are exchanged into the resultant binary. Therefore, the
stellar evolution timescale of this new binary is shorter than the
original one. Even if the new binary does produce a blue straggler, it
will do so sooner and the blue straggler is likely to have evolved
away to the giant branch before the present day
\cite{2006A&A...459..489D}.

The major effect of the cluster environment is to blur the distinction
between the two formation mechanisms for blue stragglers. Direct
collisions are much more likely between two {\it binary} stars rather
than two single stars, since the cross section of a binary star goes
as its semi-major axis instead of the stellar radius. During resonant
interactions, two or more of the stars involved could run into each
other. Even though the velocities are larger than in the single star
case (because of the addition of the orbital velocities of the
binaries), the collisions are still sufficiently gentle that our
models of collision products still apply. Therefore, identifying an
individual blue straggler as a collision product or a binary product
may not be an easy, or even a valid, distinction.

\section{Which Mechanism Dominates?}

Standard practice in science, when presented with two likely
explanations for any phenomenon, is to figure out which explanation is
correct. In astronomy, observations are made in many wavelengths and
modes; theorists produce detailed models of individual objects and of
whole populations. Comparisons are made, and after a while, one
explanation emerges triumphant over the other. However, in the
question of blue straggler formation, we are not yet at the final
stage of this story. In fact, it seems that the question we need to
answer is not ``which mechanism dominates'' but rather ``does one
mechanism dominate?'' And, in fact, we many need to look for different
dominant mechanisms in different environments in the universe and
perhaps even in different regions in the same star cluster.

As we have discussed above, both collisions and mass transfer can
reproduce observed properties of blue stragglers. These mechanisms
create main sequence stars, more massive than the turnoff, with
lifetimes that are long enough that we expect to see significant
numbers of these objects in present-day clusters. Therefore, in order
to distinguish between the different formation mechanisms, we must
look at the entire population of blue stragglers in clusters. Can either
mechanism correctly predict the distribution of blue stragglers in the
cluster, and in the colour-magnitude diagram? Which mechanism can
correctly predict the number of blue stragglers found in each cluster,
or predict the sizes of populations as a function of global cluster
properties? Thanks to both increasingly detailed observations of blue
stragglers, and to more and more models, we are starting to be able to
answer these questions.

An early comparison of collisional blue straggler populations to
observations of cluster populations was done for 6 globular clusters
\cite{2003ApJ...588..464F} with a range of central densities. The
authors compared the luminosity functions of blue stragglers with
collisional models, as well as their distribution in
temperature. Because these were HST observations, only core blue
stragglers were included in the comparison. The collisional models
could fit the data reasonably well for most clusters, but not for the
lowest-density cluster in the sample (NGC 288). This was expected
since the collision rate in that cluster is quite low. Most of the
clusters have a population of blue stragglers which are redder than
the collisional models can explain, as it is very difficult to get a
star to remain in the Hertzsprung gap for any substantial length of
time.

A similar study of the blue straggler population throughout the nearby
cluster 47 Tucanae \cite{2006ApJ...650..195M} made use of HST and
ground-based data, thereby covering the entire extent of the blue
straggler population in this cluster. In this case, the blue straggler
population was assumed to be collisional, and the models allowed for
two free parameters: the mass function of the main sequence stars
which went into the collisions, and the duration over which the
collisions could occur. The collisional models fit the observations
very well. Interestingly, the blue straggler models implied that the
mass function in the core of the cluster is very heavily weighted
towards high mass stars, and that it becomes shallower towards the
cluster edge. This result is consistent with the direct measurement of
the main sequence mass function as a function of radius in 47
Tuc. Therefore, using blue stragglers as tracers of the dynamical
state of the cluster seems plausible, which is really only possible if
collisions contribute significantly to the formation of these objects.

However, around the same time, some completely contradictory evidence
was published. By comparing the fraction of blue stragglers in
clusters to the current collision rate, Piotto et
al. \cite{2004ApJ...604L.109P} showed that there was not the expected
correlation. In fact, there seemed to be an {\em
  anti}-correlation. That paper concluded that collisions could not be
the dominant mechanism for producing blue stragglers. A consistent
selection of core blue stragglers, and comparison to many cluster
properties (e.g. velocity dispersion, central density, etc), concurred
with that result \cite{2007ApJ...661..210L,
  2008ApJ...678..564L}. Finally, a robust correlation with a cluster
property was found \cite{2009Natur.457..288K}: The number of core blue
stragglers is proportional to the cluster's core mass. The authors
conclusively proved that the number of blue stragglers could not be
proportional to the collision rate. In addition, they showed that the
blue straggler number was not linearly proportional to the core mass,
but instead goes as $\sim M_{c}^{0.5}$. If the binary fraction in
globular clusters also depends on the (core) mass as $\sim
M_{c}^{-0.5}$, then it seems reasonable that the number of blue
stragglers in globular cluster cores is proportional to the number of
binaries in the core, $f_b M_c$. In fact, preliminary evidence does
show approximately this dependence of binary fraction on cluster mass
\cite{2007MNRAS.380..781S, 2008MmSAI..79..623M}. In addition, a
correlation between blue straggler populations and binary fraction was
found for some sparse globular clusters
\cite{2008A&A...481..701S}. Therefore, binary stars must be involved
in the formation of blue stragglers, even in the cores of clusters.

Additional evidence supporting a binary evolution for blue stragglers
comes from detailed observations of surface chemical abundances in
these stars.  Ferraro {\it et al.} \cite{2006ApJ...647L..53F} did a spectroscopic survey
of blue stragglers in 47 Tuc, and found some evidence for mass
transfer -- a subset of the blue stragglers were depleted in both
carbon and oxygen compared to normal stars in the cluster. One way to
do this is to have CNO-processed material from the inner regions of a
star be dumped onto the surface of a companion, where it is now
visible. Unfortunately, a similar study of blue stragglers in M4
\cite{2010ApJ...719L.121L} found no evidence of such a population so
this signature seems not to be ubiquitous.

So, collisions can explain the observations of blue stragglers and can
predict cluster properties....except that blue stragglers cannot be
formed through collisions and must have a binary origin. The picture
is not becoming clearer. And, indeed, we have evidence that {\em both}
mechanisms are going on in the same cluster. First, we have the
long-known ``bimodal distribution'' of blue stragglers in globular
clusters. HST plus ground-based observations of a number of clusters,
starting with M3 \cite{1997A&A...324..915F}, have shown that the blue
straggler population is largest in the centre of the cluster. Then
there is a significant drop of blue straggler numbers with increasing
cluster radius, and after some point the numbers go back up again,
although not usually to values as high as the central value. The best
explanation for this distribution was presented by Mapelli et
al. \cite{2006MNRAS.373..361M}. A blue straggler population which
consists of a collisional population and a binary population was
allowed to react dynamically with the underlying cluster stars for the
typical lifetime of a blue straggler. The collisional population was
initially found only in the core of the cluster, where the binary
population was initially spread evenly throughout the cluster. These
binary blue stragglers started to fall into the core, because of mass
segregation. But because of the finite lifetime of blue stragglers,
not all of them have a chance to reach the core in the lifetime of the
cluster. The radius of the minimum of the blue straggler population is
called the ``zone of avoidance'', and is where the dynamical friction
timescale is short enough that the blue stragglers can have reached
the core. Outside of that radius, the blue stragglers remain
more-or-less at their initial radii. As shown in Mapelli {\it et al.}
\cite{2006MNRAS.373..361M} and later papers
\cite{2007ApJ...663.1040L,2008ApJ...679..712B}, the predicted zone of
avoidance matches the minimum blue straggler position within a factor
of 2 for most clusters.

\begin{figure}
  \includegraphics[height=.5\textheight]{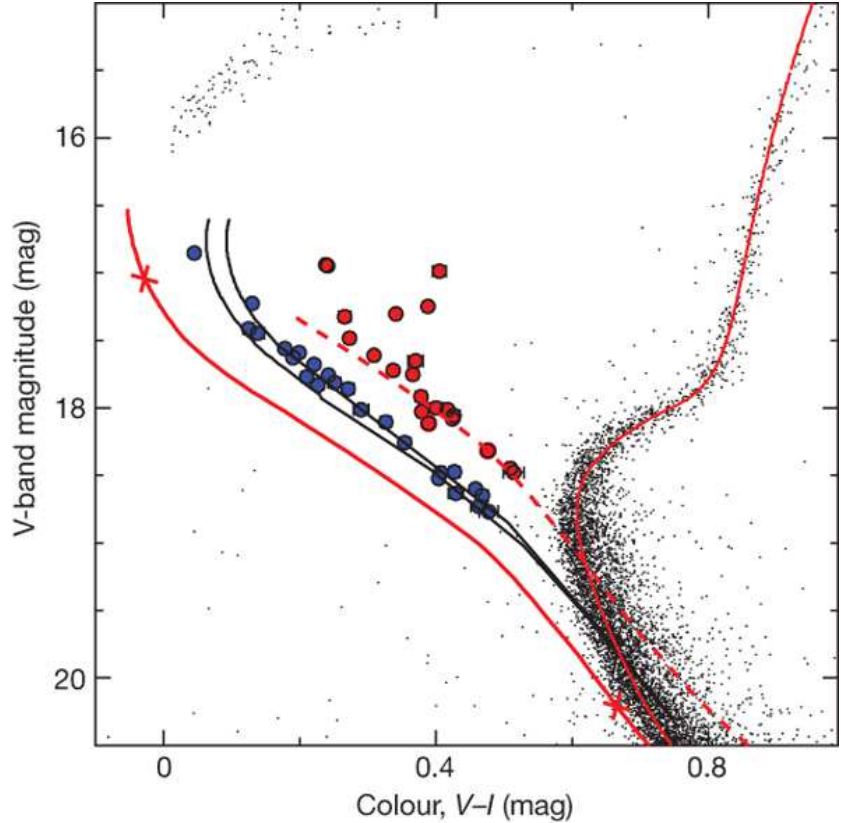}
  \caption{Colour-magnitude diagram of M30, from HST WF/PC2
    photometry. The two sequences of blue stragglers are shown as red
    and blue dots. The solid red lines are a 0.5 Gyr isochrone
    (equivalent to the ZAMS) and a 13 Gyr isochrone, which is
    appropriate for this cluster. The black lines show collisional
    isochrones, 1 and 2 Gyr after the collision events occurred. The
    dashed red line is 0.75 magnitudes brighter than the ZAMS. Figure
    taken from Ferraro {\it et al.} \cite{2009Natur.462.1028F}}
  \label{FerraroNature}
\end{figure}

The second strong piece of evidence for two formation mechanisms at work in the
same cluster comes from HST observations of M30
\cite{2009Natur.462.1028F}. In the core of this cluster, two distinct
sequences of blue stragglers can be seen (see figure
\ref{FerraroNature}). The bluer sequence lies slightly above the ZAMS,
with the brightest of these blue stragglers further from the ZAMS than
the fainter stars on this sequence. It is extremely well-modeled with
an isochrone of collision products, approximately 1 Gyr after the
collisions have occurred. The red sequence consists of stars which are
at least 0.75 magnitudes brighter than the ZAMS, and some are
substantially brighter than that and lie in the Hertzsprung gap. These
stars are consistent with the mass transfer models from
\cite{2008MNRAS.384.1263C}. The redder blue stragglers are more
centrally concentrated than the blue ones, which has interesting
implications for the binary properties of the two progenitor
populations. In another Gyr or so, normal stellar evolution will blur
the distinction between the two sequences. The fact that we see two
sequences at the present time suggests that the formation of this
population of blue stragglers was both recent and short-lived. It is
interesting to consider that this may be an indication of core
collapse or some other clearly-defined dynamical event.

\begin{figure}
  \includegraphics[height=.5\textheight]{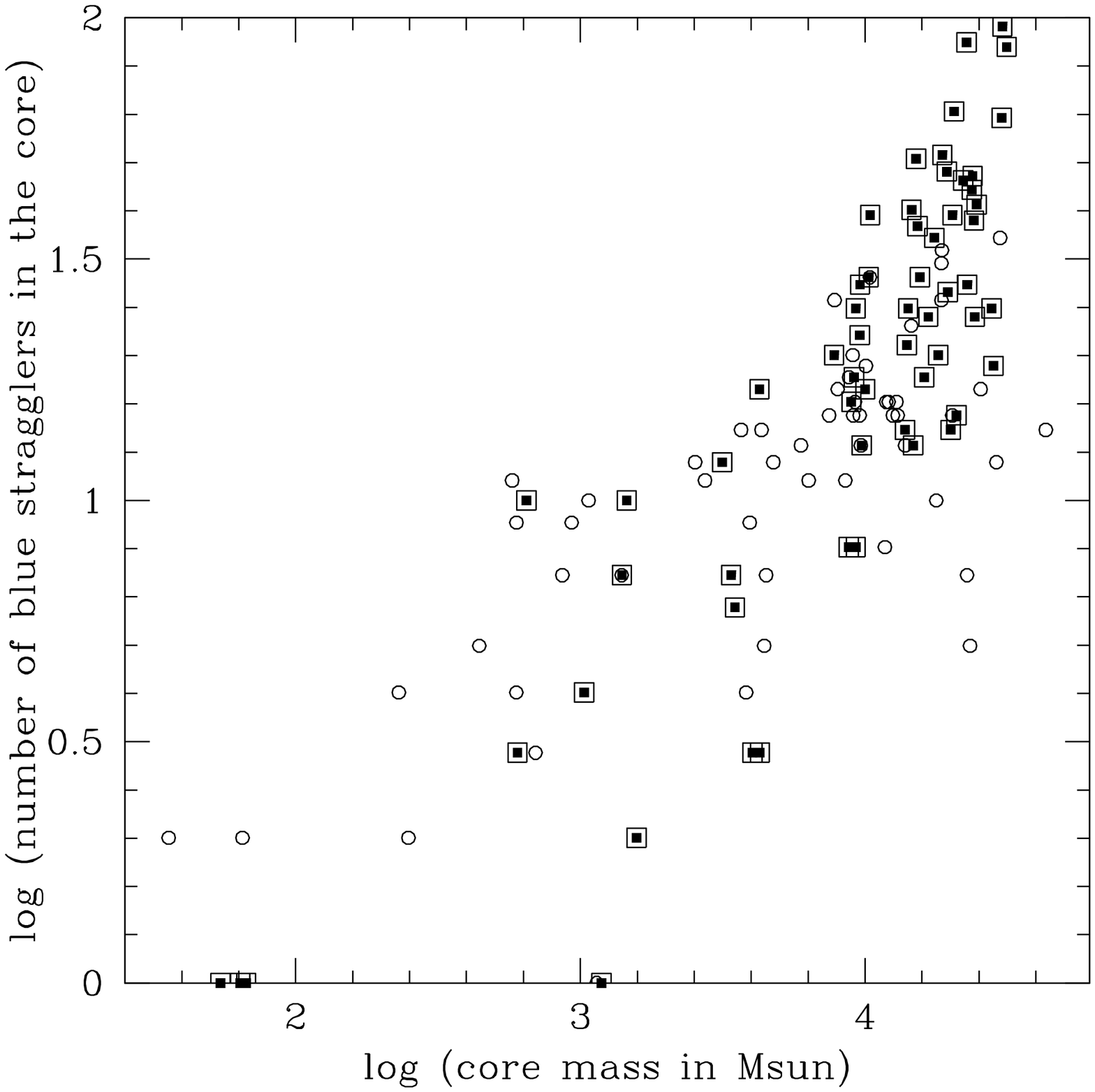}
  \caption{Number of blue stragglers in the core of globular clusters,
    vs. core mass. Observations are shown as open circles, model
    calculations are dark squares. The models require a binary
    fraction of 5-10\%, and contributions from both collisions and
    binary evolution, to match the observations.}
  \label{chatterjee}
\end{figure}

Finally, evidence that we must include both (or hybrid) formation
mechanisms comes from ``kitchen-sink'' dynamical models of globular
cluster evolution. This result was first presented in the context of
N-body models of the old open cluster M67 \cite{2005MNRAS.363..293H},
where the blue stragglers were found to have very complicated
formation histories involving both dynamics and binary
evolution. Recently, we have been investigating the blue straggler
populations of approximately 100 globular cluster models using the
Northwestern Monte Carlo code (Sills, Chatterjee \& Glebbeek, in
preparation). Our preliminary results are shown in figure
\ref{chatterjee}. Here we show the observed trend of core blue
straggler number with core mass (open circles), and the results of our
simulations overplotted (squares). These models, which reproduce the
observations very nicely, require a cluster binary fraction of
5-10\%. Cluster with no initial binaries simply do not produce enough
blue stragglers, especially in the high-mass clusters. We can also
look at the formation histories of all our modeled blue stragglers,
and we find that very few of them are purely collision products or
purely binary mass transfer products. Almost all of them have both 
dynamical encounters and binary evolution in their history. We can also
distinguish between blue stragglers formed in a binary encounter from
those formed from direct collision, and we find that the binary
encounters significantly outnumber both direct collisions and binary
mass transfer.

\section{Future Work and Open Questions} 

We are converging on an explanation for the formation of blue
stragglers in clusters. Unfortunately it seems to be a complicated
explanation, involving multiple formation mechanisms which act with
different weights in different environments and at different times in
the cluster evolution. However, we are certainly making progress with
a combination of detailed observations and more sophisticated models
of both individual blue stragglers and cluster populations. As always,
there are still a number of additional complications and
considerations that we still need to consider. The following is a
combination of a ``wish list'' and a nod to additional physics we may
still need to include in our models.

For the most part, we have completely ignored the effects of any
system more complicated than binary stars.  However, there has been
some promising research recently pointing to the importance of triple
systems, particularly for blue straggler populations in old open
clusters \cite{2009ApJ...697.1048P, 2010arXiv1009.0461L}.  Modeling the
dynamical effects of triples is extremely computational intensive in
N-body codes, and they are completely neglected in currently Monte
Carlo codes, but we may have become cleverer if we really wish to
investigate all possible formation mechanisms of blue stragglers and
other stellar exotica in clusters in the context of full-blown cluster
models.

In the mid-90's, there was a concerted effort to understand
single-single and single-binary encounters. Calculations of encounter
outcomes and cross sections were automated and parameterized, and so
we have a fairly good understanding of these kinds of
encounters. However, because the parameter space of binary-binary
encounters was so large, and because it is much harder to determine
when an encounter is ``finished'', we know much less about the
outcomes of higher order encounters. And, of course, very little work
has been done on triples and higher order multiples, as discussed
above. Special-purpose codes such as FEWBODY
\cite{2004MNRAS.352....1F} have been built to follow encounters of a
small number of stars, and could be harnessed for this purpose. Brute
force, however, is unlikely to cover the interesting parameter space
in a useful amount of time, and so the problem of being selective in
initial conditions is likely to remain.

I have not mentioned the rotation rates of blue stragglers in
clusters, and the information that those measurements can give
us. There are very few measured rotation rates of blue stragglers,
particularly in globular clusters. The situation is rather better in
open clusters, where the fields are sparser and spectroscopic studies
have been going on for longer (e.g. for characterization of the binary
populations in those clusters\cite{2009Natur.462.1032M}). The
indications are that blue stragglers do not rotate extremely rapidly,
but that they rotate more rapidly than normal stars of the same
temperature (typically 10-40 km/s for blue stragglers, compared to
2-10 km/s for normal stars). Have they been spun up by collisions or
mass transfer? Have they had a chance to spin down from a very
significant rotation rate? Can the rotation rates tell us anything
about their time of formation? Does the rotation mix the stars
significantly, so that their evolutionary tracks are different from
non-rotating stars?  Models of the angular momentum evolution of both
collision products e.g. \cite{2005MNRAS.358..716S} and binary mass
transfer products are required before we can use these pieces of
observational evidence to understand anything more about blue
straggler formation in clusters.

We have detailed evolutionary models of some binary mass transfer
products. However, the (very large) parameter space has not yet been
sufficiently explored. We need to understand enough about the
evolution of mass transfer products that we can approximate their
evolution in binary evolution codes like BSE
\cite{2002MNRAS.329..897H}. This has been done for collision products,
to some extent -- the Make Me A Star code (MMAS)
\cite{2002ApJ...568..939L} will produce a collision product from two
parents, without the need for the detailed hydrodynamic
calculation. The subsequent evolution of this collision product can be
calculated using a stellar evolution code, or the lifetime and CMD
location can be estimated based on the core hydrogen fraction of the
product. We are not yet at this stage for binary mass transfer models.

And finally, it would be of great interest to a large population of
researchers to know the binary properties in globular clusters. Does
the binary fraction depend on the mass of the cluster, as suggested by
preliminary work? Does it depend on the location in the cluster? The
answer to this question seems to be ``yes'', based on some scant
observations and a number of dynamical models. Does it depend on which
stellar population one observes? Of particular interest to
understanding the formation of blue stragglers, we're looking for two
answers to this question. One, what is the binary fraction of main
sequence-main sequence binaries, which could form blue stragglers
later? Two, what is the binary fraction of the blue stragglers
themselves? And, of course, any information we can glean about the
mass ratio, semi-major axis, and eccentricity distributions of any of
these binaries will be crucial for many of us. Will there be a
multi-object spectrograph on a telescope with sufficient resolution to
disentangle the crowded conditions of globular clusters any time soon?
I know of no such instrument in the planning stages, so we may be
using approximations and indirect methods for a long time.

\bibliographystyle{aipprocl} 

\end{document}